\documentclass[12pt]{iopart}
\usepackage[dvips]{color}
\usepackage{graphicx}

\def\lsim{\mathrel{\raise2pt\hbox to 8pt{\raise -5pt\hbox{$\sim$}\hss{$<$}}}}

\def\pmb#1{\setbox0=\hbox{#1}
\kern-.025em\copy0\kern-\wd0
\kern.05em\copy0\kern-\wd0
\kern-.025em\raise.0433em\box0}

\def\half{{\textstyle {1\over2}}}
\def\beq{\begin{equation}}
\def\eeq{\end{equation}}
\def\beqa{\begin{eqnarray}}
\def\eeqa{\end{eqnarray}}

\def\ttimes{{\scriptstyle \times}}

\begin{document}

\title[Metastable strange matter and compact quark stars]{Metastable strange matter and compact quark stars}

\author{M. Malheiro\dag\ \footnote[2]{Email: mane@if.uff.br},  M. Fiolhais\S\ and A. R. Taurines\P\
}

\address{\dag\ Instituto de F\'\i sica, Universidade Federal Fluminense,
24210-340 Niter\'oi, Brazil}

\address{\S\ Departamento de F\'\i sica and Centro de F\'\i sica Computacional,\\
Universidade de Coimbra, P-3004-516 Coimbra, Portugal}

\address{\P\ Instituto de F\'\i sica, Universidade Federal do Rio Grande do Sul
CP 15051, 91501-970 Porto Alegre, Brazil}

\begin{abstract}
Strange quark matter in beta equilibrium at high densities is studied in a
quark confinement model. Two equations of state are dynamically generated
for the {\it same} set of model parameters used to describe the nucleon:
one corresponds to a chiral restored phase with almost massless quarks and
the other to a chiral broken phase. The chiral symmetric phase saturates at around five
times the nuclear matter density. Using the equation of state for
this phase, compact
bare quark stars are obtained with radii and
masses in the ranges $R\sim 5 - 8$~km and $M\sim M_\odot$. The
energy per baryon number decreases very slowly from the center of the star to the periphery, remaining above
the corresponding values for the iron or the nuclear matter, even at the edge.
Our results point out that strange quark matter at very high densities may
not be absolutely stable and the existence of an energy barrier
between the two phases may prevent the compact quarks stars to decay to
hybrid stars.  

\end{abstract}


\submitto{\JPG}

\maketitle

\section{Introduction}

The recent discoveries of X-ray sources, whose origin may be attributed to strange
stars~\cite{bombaci,li,dey}, is stimulating
theoretical research on the structure, composition, dynamics and evolution of such objects.
It is believed that the innermost region of compact stars, such as neutron stars,
can be made out of hyperons, meson condensates or 
quark matter~\cite{glendenning,glendenning1,fridolin,glendenning2,heiselberg},
whose thermodynamical behavior can be expressed by means
of a relativistic equation of state (EOS)~\cite{glendenning,fridolin,taurines1}.

Various effective models for
the nucleon, using quarks as fundamental fields, have been used to provide
EOS's for quark matter, which are subsequently  applied to investigate the structure
of compact stars~\cite{dey,heiselberg,greiner,witten}.
In this vein, we use the chromodielectric model (CDM)~\cite{pirne1,manoj} to obtain
EOS's for dense quark matter, and
report, in this letter,
the properties of the resulting quark stars. The chiral CDM provides a reasonable
phenomenology for the nucleon~\cite{neuber,drago1} and N--N$^*$ transition amplitudes
\cite{ndelta1}. Baryons appear as solitons with three quarks
dynamically confined by a scalar field, $\chi$, whose quanta can be assigned
 to $0^{++}$ glueballs. In the quark matter sector, the model  yields a relatively soft
EOS at large densities~\cite{drago2,rosina}. Drago et al.
applied the CDM with a quadratic potential to the structure of compact stars~\cite{drago,drago3}.
They connected the EOS for quark matter, as provided
 by the CDM, with an EOS for hadron matter, obtaining stars with masses in the range
$1-2 M_\odot$ and radii of the order 10 km with a hadron crust of 2 km.

In the present work we consider an extension of the model used in Ref.~\cite{drago},
now taking quartic instead of quadratic potentials.
For such potentials two qualitatively different EOS's are dynamically generated for the
{\em same} set of model parameters (the same used to describe the nucleon).
They correspond to two distinct phases: in one phase the $\chi$ field is large and  the
quarks are massless (chiral symmetric phase). In the other one, $\chi$ is small and quarks
are massive (chiral broken phase). The EOS's are degenerate at very high densities but
with an energy barrier between them. As the density decreases, they separate and saturate
at very different densities, yielding different energies per baryon number.
The EOS for the chiral symmetric matter saturates at densities much higher than the
nuclear matter
equilibrium density and with an energy per baryon number above that quantity in iron or
in nuclear matter at low densities.
The compact stars obtained using this EOS
are made out of quarks only, since the density at the edge is much above the
nuclear matter saturation density, and hadronization processes do not take place.

The prediction of smaller and denser objects in comparison with the neutron
stars is quite exciting in view of
the recent discovery of
X-ray sources, by the Hubble and Chandra telescopes,
which increased the plausibility that these sources might be
strange quark stars~\cite{bombaci,li,dey}.
In particular,
the isolated compact object RX J1856.5-3574, not showing evidence
for spectral lines or edge features~\cite{pons,drake},
reinforced the conjecture for the existence of
strange matter stars. However, the first results indicating a small radius for that compact object 
were based on the reported parallax of Ref.~\cite{walter1}, which was not taking 
into account the camera geometrical distortion. Considering these corrections 
the parallax was reduced by roughly a factor of two ~\cite{walter,kaplan} 
and the distance has been revised from $60$ to $117$ pc approximately.
The revised radius ($R = 11.4\pm 2 $~km) and mass ($M/M_\odot = 1.7 \pm 0.4$) make that 
isolated compact object reproduceable by many EOS with and without strange matter ~\cite{walter,prakash}. 
Hence, it is not guaranteed that RX~J1856.5-3574 might only be described as a quark star.

\section{Quark matter}

The CDM Lagrangian can be written as~\cite{pirne1,manoj,neuber}
\begin{equation}
  \mathcal{L} = \mathcal{L}_q + \mathcal{L}_{\sigma,\pi}
    +  \mathcal{L}_{q-\mathrm{meson}} +  \mathcal{L}_\chi\;,
  \label{langrangian}
\end{equation}
where
\begin{eqnarray}
  \mathcal{L}_q &=& \mathrm{i}\bar{\psi}\gamma^\mu \partial_\mu\psi \;,
\nonumber \\
  \mathcal{L}_{\sigma,\pi} &=&
  \half\partial_\mu\hat{\sigma}\partial^\mu\hat{\sigma}
  + \half\partial_\mu\hat{\vec{\pi}}\cdot\partial^\mu\hat{\vec{\pi}}
  - {W}(\hat{\vec{\pi}},\hat{\sigma})\;,
  \label{langrangian1}
\end{eqnarray}
${W}(\hat{\vec{\pi}},\hat{\sigma})$ being the usual
Mexican hat potential for the chiral mesons sigma and pi. The quark-meson interaction, assuming two flavours, is
\begin{equation}
    \mathcal{L}_{q-{\mathrm{meson}}} = {g\over\chi}\, \bar{\psi}
    (\hat{\sigma}+\mathrm{i}\vec{\tau}\cdot\hat{\vec{\pi}}\gamma_5)
    \psi\;.
   \label{langrangian2}
\end{equation}
The last term in (\ref{langrangian})
contains the kinetic and the potential piece for the $\chi$-field:
\begin{equation}
  \mathcal{L}_\chi =
  \half\partial_\mu\hat{\chi}\,\partial^\mu\hat{\chi}
  -  U(\hat{\chi})\, .
  \label{langrangian3}
\end{equation}
We consider a quartic potential
\begin{equation}
U(\chi)= {1 \over 2} M^2 \chi^2
         \left[  1+\left( {8 \eta^4 \over \gamma^2}-2 \right) {\chi \over \gamma M} +
\left( 1- {6 \eta^4 \over \gamma^2} \right) {\chi^2 \over (\gamma M)^2} \right], \,
\ \ \ \ \label{uchi}
\end{equation}
where $M$ is the $\chi$ mass. It has a global minimum at $\chi=0$
and a local one at $\chi=\gamma M$.
The height of the local minimum, $B=U(\gamma M)=(\eta M)^4$, may be interpreted
as a ``bag pressure"~\cite{rosina2}, as in the MIT bag model, and that will be
used to fix parameters in $U(\chi)$.

In the soliton sector of the model, a good description of the nucleon is obtained
for $\chi$ close to zero, a region of the potential (\ref{uchi}) where the cubic
and quartic terms play no role.
All dependences go into just
one parameter, namely $G=\sqrt{g M}$, and best nucleon properties are obtained for
$G\sim 0.2$ GeV (e.g. $M=1.7$~GeV and $g=0.023$~GeV~\cite{drago1}).
In non-strange homogeneous quark matter the deppendence on the single parameter $G$ is even
exact for the {\em quadratic} model \cite{mcgovern}. In our study we keep that value for $G$.
In the quark matter sector of the model with double minimum potential
there are solutions with small $\chi$ and solutions with large $\chi$, i.e.
in the region of the local minimum of the potential, as discussed below in more detail.
From $B=\eta^4 M^4$ and assuming  the wide range $0.150 \le B^{1/4} \leq 0.250$ GeV,
one has $0.08\le \eta \le 0.15$, using $M=1.7$~GeV. The parameter $\gamma$,
which does not affect
the phenomenology of the homogeneous matter, is not
 a totally free parameter: $\gamma^2\ge 6 \eta^4$  since the quartic term in (\ref{uchi})
must be positive.

The extension of the  model to include the strange quark, requires that one more
term be added to the interaction Lagrangian (\ref{langrangian2}),
accounting for
the coupling between the strange quark and the  $\chi$ field. We take it in the
simplest form~\cite{birsegovern}
\begin{equation}
    \mathcal{L}_{s-{\mathrm{meson}}} = {g_s \over\chi}\, \bar{\psi}_s
    \psi_s\;.
   \label{langrangian4}
\end{equation}
In addition, for the sake of beta equilibrium, an electron gas must also be included.
The mean-field energy per unit volume for a homogeneous system of $u$,
$d$ and $s$ quarks, interacting with $\chi$ and $\sigma$, plus electrons, is
\begin{eqnarray}
&&\varepsilon = \alpha
\sum_{f=u,d}\int_{0}^{k_{f}}\frac{d^3k}{(2\pi)^3}\sqrt{k^2+m_f(\sigma,\chi)^2}\nonumber \\
& &+ \alpha
\int_{0}^{k_{s}}\frac{d^3k}{(2\pi)^3}\sqrt{k^2+m_s(\chi)^2}
 + 2
\int_{0}^{k_{e}}\frac{d^3k}{(2\pi)^3}\sqrt{k^2+m_e^2} \nonumber \\
& &+{ U}(\chi)+{ m _\sigma^2 \over 8 \, f_\pi^2}(\sigma^2-f_\pi^2)^2\, ,
\label{eov2}
\end{eqnarray}
[$f_\pi=93$~MeV and we always use $m_\sigma=1.2$~GeV in this paper, though the results
are not sensitive to the sigma mass].
The first two terms refer to quarks, and the third one to the electrons,
all described by plane waves. The degeneracy factor is
 $\alpha=6$ (for spin and color).
The Fermi momentum for each type of particle, $k_i$, is related to the corresponding
density, $\rho_i$, through $\rho_i={\alpha  k_i^3 / (6 \pi^2)}$.
The quark masses in (\ref{eov2}) are all different \cite{birsegovern}:
\begin{eqnarray}
m_{u,d} = \frac{g_{u,d}\, \sigma}{\chi \, f_\pi}, \,\,\, \,\,\,\,\, m_{s} =
\frac{g_{s}}{\chi}\, ,
\label{qmass}
\end{eqnarray}
with the coupling constants for each flavour given by
$g_u= g \, (f_\pi+\xi_3)$, $g_d= g \, (f_\pi-\xi_3)$ and $g_s=g\, (2 f_k-f_\pi)$
[$\xi_3=-0.75$ MeV, $f_K=113$ MeV].
Since the vacuum expectation value of the confining field is zero,
the quark masses raise up to infinity for densities approaching zero.

A variational principle applied to the energy density, Eq.~(\ref{eov2}),
leads to two gap equations for $\sigma$ and $\chi$.
In the interior of a compact star the matter should
satisfy both the electrical charge
neutrality and chemical equilibrium.
These conditions together with the gap equations,
form a system
of six algebraic equations solved at each baryon density
$
\rho={1 \over 3} \left( \rho_u+\rho_d+\rho_s \right)$.
For a given $\rho$ one obtains a self-consistent set
$\sigma$, $\chi$, $k_u$, $k_d$,  $k_s$, $k_e$ and,
with such solution for each density, the energy density [Eq. (\ref{eov2})] is readily evaluated
as well as the energy per baryon number, $\epsilon/\rho$, the pressure, etc.

For the same set of
parameters we found two distinct stable solutions, hereafter denoted by I and II.
For both solutions, $\sigma$ remains always close to $f_\pi$ irrespective of the density.
In solution I, the $\chi$ field, close to zero, is a slowly increasing function of
 the density. For small
$\chi$, the quartic potential (\ref{uchi}) is indistinguishable from
$U={1\over 2}M^2 \chi^2$, thus, in practice, our solution
 I is the one obtained and used by Drago et al.
\cite{drago} in the framework of the quadratic potential. Due to the smallness of the
$\chi$ field, quark masses are large [see Eq.~(\ref{qmass})] and the
system is in a chiral broken phase.
On the other hand, for solution II, the confining field is large, $\chi\sim \gamma M$
(local minimum of $U$), almost independent of the density.
The resulting quark masses are similar for the three flavours and very close to zero
(chiral restored phase).
Because the chemical potentials  are dominated by the Fermi momentum,
one has $\mu_u\simeq\mu_d\simeq\mu_s$, and
therefore $\mu_e\simeq0$, i.e. in solution II there are
almost no electrons.
Besides solutions I and II, there is an additional
{\em unstable} solution corresponding to $\chi\sim \gamma M /2$
[local maximum of $U(\chi$)].

\begin{figure}[bht]
\centerline{\includegraphics[clip=on,width=13cm]{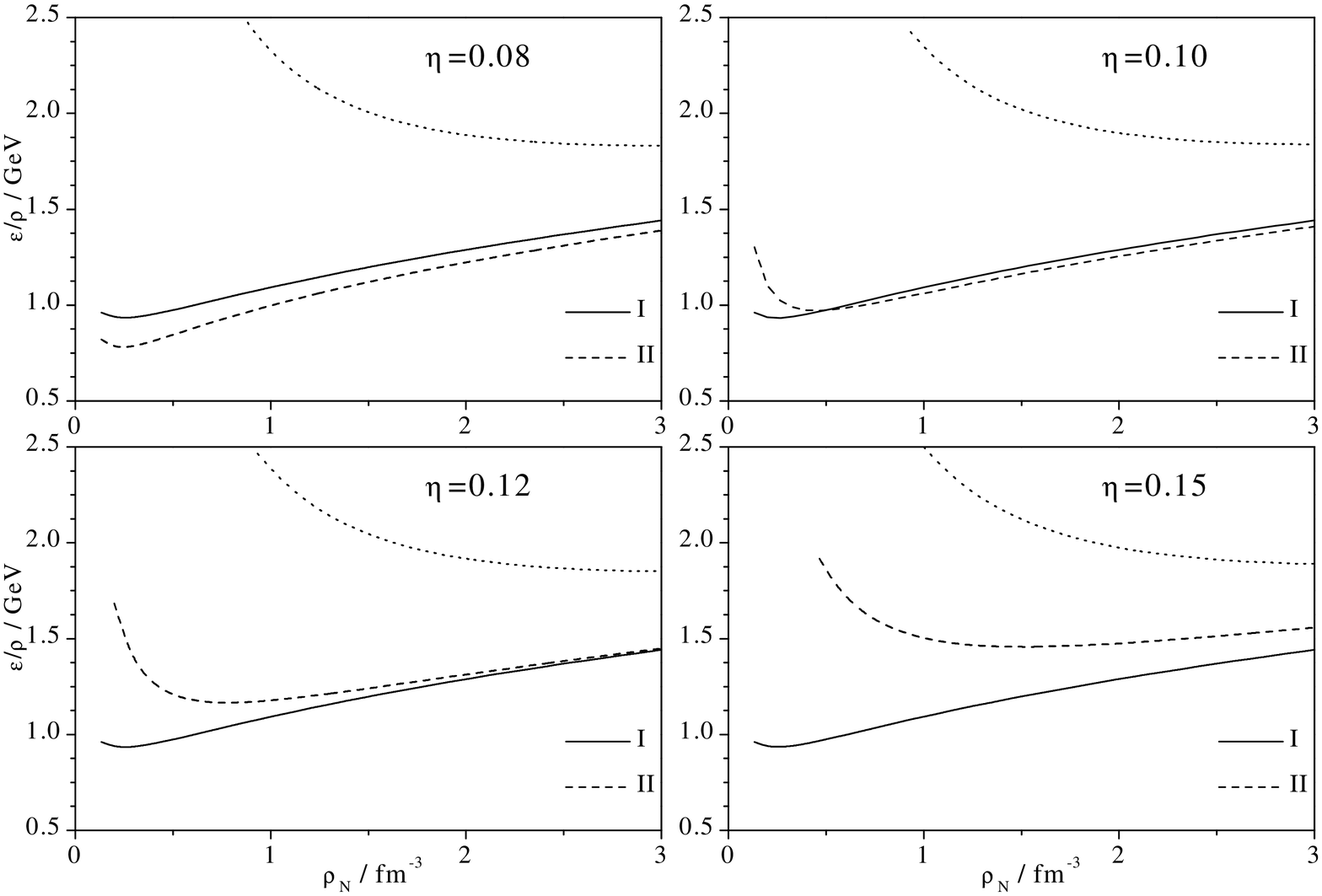}}
\caption{Energy per baryon number versus density for solutions I (solid line, small $\chi$)
and II (dashed line, $\chi\sim \gamma M$) and various parameters $\eta$ (other model parameters:
$M=1.7$~GeV, $g=0.023$~GeV and $\gamma=0.2$). The dotted line corresponds to the
unstable solution with $\chi \sim \gamma M /2$.}
\label {fig1}
\end{figure}

For each solution we obtained the corresponding energy per baryon number as a
function of the baryon density (EOS)  (see Fig.~\ref{fig1}).
EOS-I is not sensitive to $\gamma$ and $\eta$ (since $\chi$ is small), just depends on
$G$, and it is rather similar to the one used in Ref.~\cite{drago}.
The saturation density occurs at a low density, slightly higher than the nuclear
matter equilibrium density, $\rho_0$.  Its shape, at intermediate densities, 
is similar to hadronic EOS's (see Ref.~\cite{drago2} for the two flavours sector).

\begin{figure}[bht]
\centerline{\includegraphics[clip=on,width=12cm]{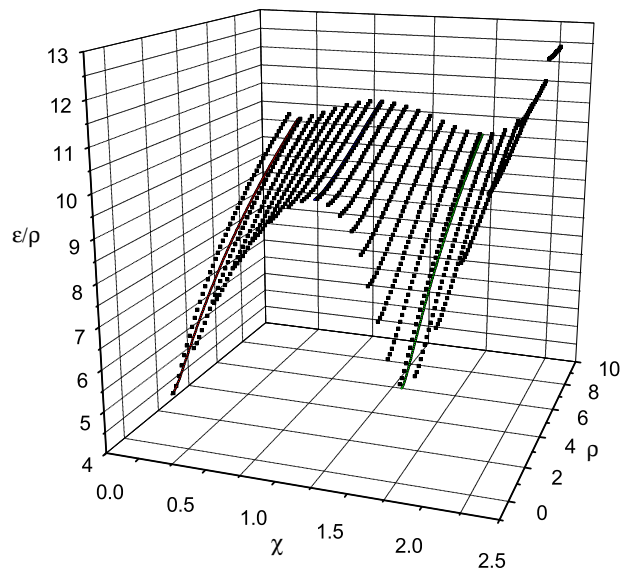}}
\caption{ 3D plot of the energy per baryon number versus density and $\chi$ field. Solutions I lies
in the  valley for small $\chi$, the unstable solution is at the maximum and solution II 
lies in the valey at high $\chi$. The parameters are as in Fig.~\ref{fig1}.} 
\label {fig2}
\end{figure}

 The EOS-II is also insensitive to $\gamma$, but does depend on $\eta$ [in fact, the
dependence is on $(\eta M)^4$, as we have already discussed]: the energy per baryon
number increases with $\eta$ and so does the saturation density.
Depending on $\eta$ the minimal energy per baryon number of solution II can be either
below or above solution I. For $\eta\sim 0.12$ the saturation occurs at $\rho\sim 5 \rho_0$
and the energy per baryon number is some 230 MeV
higher than for solution I at its saturation density. The two stable solutions are almost
degenerated at high densities in the narrow range  $0.1 \le \eta \le 0.12$.
In Fig.~\ref{fig1} the dotted lines refer to the EOS for the unstabe solution of the
gap equations (supplemented by electric neutrality and beta equilibrium conditions), for which
the $\chi$ is at the local maximum of the potential (\ref{uchi}).
In order to undergo a transition from I to II, the system has to go through the
 energy barrier represented by the dotted EOS. The barrier gets higher
at small densities and, for $\eta=0.12$ (third panel in Fig.\ref{fig1}), at
$\rho\sim 5 \rho_0$, $\epsilon/\rho\sim 2.7$~GeV for the unstable solution. Therefore
a transition from one regime to the other is not likelly to occur and both minima in the
EOS I and II  are stable. In a 3D plot of the energy per baryon number versus $(\rho,\chi)$
the stable solutions correspond to two distinct
``valleys", and the unstable solution  corresponds to the top of
 the barrier between these two valleys as it is shown in Fig.~\ref{fig2}.
Our results point out that the metastability of
strange quark matter only occurs at high densities for solution II.

Interesting enough, our results with two distinct EOS, are consistent with the
results from a recent calculation in perturbative QCD~\cite{fraga}. We checked that,
for $\eta\sim0.12$ the CDM reproduces accurately the EOS's obtained in Ref.~\cite{fraga}.
However, in our case, there is no parameter fit to get two equations of state: they are
dynamically generated for the same Lagrangian parameters.
Also in Ref.~\cite{peshier}, using an extension to finite chemical
potential of lattice QCD data for the equation of state,
similar results were obtained. Therefore, there seems
to be some model independence, which is worth to point out here.

To investigate the structure of stars we solved the Tolman-Oppenheimer-Volkoff equation.
Since EOS-I is identical to the one using a quadratic potential, it leads to
stars that have the same phenomenology as the hybrid stars obtained by Drago et al.~\cite{drago}:
$R\sim 10 - 12$ km, a hadron crust and a mass $M\sim 1- 2 M_\odot$.

\begin{figure}[thb]
\centerline{\includegraphics[clip=on,width=9.0cm]{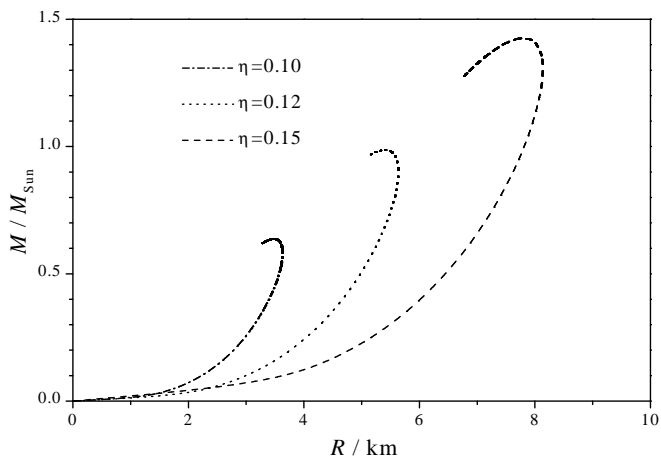}}
\vspace*{-0.3cm}
\caption{Mass versus radius for the pure quark stars (solution II) in the CDM model.
For the other model parameters
see caption of Fig.~\ref{fig1}.}
\label{fig3}
\vspace*{-0.3cm}
\end{figure}

Since EOS-II saturates at a high density and, in addition,
the system is not likely to undergo a transition to solution I,
one should not perform any connection to the hadronic sector:
the EOS-II alone generates a new family of strange quark stars.  In Fig.~\ref{fig3}
it is shown the mass-radius relation for different values of $\eta$.
These quark stars are smaller and
denser in comparison with those resulting from EOS-I. For  $\eta\sim0.115$
(and $M=1.7$~GeV, yielding $B^{1/4}\sim 0.195$ GeV) one obtains a maximum radius $R\sim 6$~km
(and a corresponding mass $M\sim 0.9 M_\odot$).
According to our calculation, such star has a central density of $10\rho_0$ ($\rho_0$
is the nuclear matter density) and a central energy density
$\epsilon\sim 3\ttimes 10^{15}$ g/cm$^3$. At the edge, the density drops to $5\rho_0$ and
$\epsilon\sim 1.35\ttimes 10^{15}$ g/cm$^3$. The ratio $\epsilon/\rho$ remains approximately constant
inside the star and the minimum period of the star, computed using the expression
given in Ref.~\cite{haensel}, is $\sim0.4$~ms.

A possible mechanism to explain the formation of such bare strange stars
could be a supercooling effect in the early universe, when a significant amount
of quark matter was frozen into quark stars. This mechanism needs an effective
potential with two minima, as discussed in~\cite{cottingham}, resembling the
quartic $\chi$ potential.
As we see from Fig.~\ref{fig3}, the mass-radius relation for these strange small stars
mainly depends on the height of the local minimum of the $\chi$ potential.
Finally, it is also worth mentioning that our results for the
maximum radius and maximum mass for the stars from solution II are in agreement with
Ref.~\cite{banerjee}, where it was shown that the Chandrasekhar
limit for quarks stars depends on $B$.

Let us summarize our results. We have considered the CDM with a quartic
potential, with parameters fixed in the nucleon sector and to yield a reasonable bag constant.
Using a mean-field  variational method we obtained two solutions for homogeneous strange
matter in beta equilibrium, one similar to the already known solution for quadratic potentials
(with massive quarks) and
a new one with massless quarks. The pure quark stars emerging from the chiral symmetric
solution are small and dense compact objects. To get so a small stars,
the saturation density
should be high enough, and this is achieved with a corresponding energy per baryon
number always above the energy
per nucleon in the iron at low density or the energy per nucleon in nuclear matter
at its saturation density.

\subsection*{Acknowledgments}

This work was supported by FCT (POCTI/FEDER program), Portugal and by CNPq/ICCTI through
the Brazilian-Portuguese scientific exchange program.

\end{document}